\documentclass[aps,showpacs,twocolumn,superscriptaddress,nofootinbib,floatfix,section 10pt]{revtex4-2}
\usepackage{amsfonts,amssymb,stmaryrd,latexsym,amsmath,braket}
\usepackage{graphicx,subfigure}
\usepackage{comment}
\usepackage{times}
\usepackage{slashed}
\usepackage{bm}
\usepackage{braket}
\usepackage{appendix}
\usepackage{enumitem}
\usepackage{multirow}
\usepackage{array}
\usepackage{amsmath}
\usepackage[colorlinks=true,backref=true,linktocpage=true,citecolor=blue,urlcolor=blue,linkcolor=blue,pdfpagemode=UseOutlines]{hyperref}
\usepackage[dvipsnames]{xcolor}

\begin{document}
\title{Training neural control variates using correlated configurations
}

\author{Hyunwoo Oh}
\email{hyunwooh@umd.edu}
\affiliation{Department of Physics and Maryland Center for Fundamental Physics, University of Maryland, College Park, MD 20742 USA}

\begin{abstract}

Neural control variates (NCVs) have emerged as a powerful tool for variance reduction in Monte Carlo (MC) simulations, particularly in high-dimensional problems where traditional control variates are difficult to construct analytically. By training neural networks to learn auxiliary functions correlated with the target observable, NCVs can significantly reduce estimator variance while preserving unbiasedness. However, a critical but often overlooked aspect of NCV training is the role of autocorrelated samples generated by Markov Chain Monte Carlo (MCMC). While such samples are typically discarded for error estimation due to their statistical redundancy, they may contain useful information about the structure of the underlying probability distribution that can benefit the training process. In this work, we systematically examine the effect of using correlated configurations in training neural control variates. We demonstrate, both conceptually and numerically, that training on correlated data can improve control variate performance, especially in settings with limited computational resources. Our analysis includes empirical results from $U(1)$ gauge theory and scalar field theory, illustrating when and how autocorrelated samples enhance NCV construction. These findings provide practical guidance for the efficient use of MCMC data in training neural networks.

\end{abstract}

\date{\today}
\maketitle

\section{Introduction} \label{Sec:Introduction}

Monte Carlo (MC) methods play a central role in the numerical study of high-dimensional integrals and expectation values, particularly in fields such as statistical physics, Bayesian inference, and quantum field theory. Despite their broad applicability, MC estimators often suffer from large statistical uncertainties, especially when sampling from complex, high-dimensional, or multimodal distributions. Variance reduction techniques are therefore crucial to improving the efficiency and reliability of MC computations.

One classical approach to variance reduction is the control variates method~\cite{Kahn1953, Fieller1954}. In this method, auxiliary functions---called control variates---are introduced that are correlated with the target observable and have known expectation values. By subtracting a suitably scaled version of the control variate from the observable, one can construct an improved estimator with reduced variance while preserving the correct mean. Traditionally, effective control variates rely on analytical insight into the structure of the integrand or the underlying probability distribution, limiting their applicability in complex systems.

Recent advances in machine learning, particularly the use of neural networks as flexible function approximators, have opened up new possibilities for constructing control variates directly from data. In this approach, a neural network is trained to learn a function correlated with the observable of interest, subject to a constraint on its expected value. These neural control variates (NCVs) offer a powerful and general framework for variance reduction, applicable even in situations where little analytic knowledge of the target distribution is available. Several recent works have demonstrated the promise of NCVs in various domains, including lattice field theory~\cite{Bedaque:2023ovz}, Bayesian inference~\cite{Wan2018, Sun2021, Sun2023, Ott2023}, and machine learning~\cite{grathwohl2018, Muller2020}, along with their theoretical understanding~\cite{Belomestny2023}.

However, the choice of training data is subtle: while traditional MC workflows rely on decorrelated configurations for error estimation, correlated configurations may carry additional information that is beneficial for training of NCVs, particularly in resource-limited settings.

In this work, we address this issue systematically. We begin by reviewing the theoretical foundation of control variates and their neural extensions. We then analyze the role of autocorrelated samples in training, highlighting when and why they can be useful despite being statistically redundant for error estimation. Throughout, we provide empirical benchmarks in representative systems to support our claims.

This paper is organized as follows. In Section~\ref{Sec:CV}, we review the control variates method for variance reduction and introduce neural control variates, which leverage machine learning to construct more effective control functions. Section~\ref{Sec:Correlated} examines the role of correlated configurations generated by Markov Chain Monte Carlo and their impact on training neural control variates. In Section~\ref{Sec:Result}, we present numerical evidence supporting the arguments of Section~\ref{Sec:Correlated}, using examples from $U(1)$ gauge theory and scalar field theory. Finally, we conclude in Section~\ref{Sec:Discussion} with a summary of our findings and possible directions for future work.

\section{Control variates} \label{Sec:CV}

A control variate $f$ is an operator with expectation value is zero\footnote{More generally, if there exists an operator $O_a$ whose expectation is known to be $a$, then one can construct a control variate as $f=O_a-a$.}. Because $\langle f \rangle =0$, it can be subtracted from an observable $O$ to define an unbiased estimator:
\begin{equation}
    \tilde{O} = O -f.
\end{equation}
Although $\tilde O$ and $O$ share the same expectation value, their variances can be different:
\begin{equation}
    {\rm Var}(\tilde O) = {\rm Var}(O) - (\langle 2 f O \rangle - \langle f^2\rangle ).
\end{equation}
This expression shows that if $f$ is highly correlated with $O$, and its variance $\langle f^2\rangle$ is small, then $\tilde O$ can have significantly reduced variance while remaining unbiased. This is the central idea behind the control variates method.

In practice, it is often difficult to construct operators whose expectation values are exactly known. One systematic method for constructing control variates exploits Stein's identity~\cite{stein1972bound}, which holds for any differentiable function $g$ under the Boltzmann weight:
\begin{equation}
    \int D\phi \, \partial(g \, {\rm e}^{-S}) = 0 .
\end{equation}
Applying this identity yields a general form for a control variate:
\begin{equation}
    f = \sum_i \left( \frac{\partial g_i}{\partial \phi_i} - g_i \frac{\partial S}{\partial \phi_i} \right) \label{Eq:VCV},
\end{equation}
where $g:M^V \to \mathbb R^V $ is a vector-valued function with components $g_i$, $S$ is the action, $M$ is the domain of integration (typically the space of field configurations), and $V$ denotes the number of degrees of freedom. This formulation allows for broad flexibility in designing $f$, and as discussed in~\cite{Si2020, Lawrence:2024xsi}, it is universal: any control variates, including the perfect (or optimal) control variates $f_P=O - \langle O \rangle$, can in principle be represented using this construction.

\subsection{Neural control variates} \label{Sec:NCV}

Since the control variate $f$ is constructed from the function $g$, an explicit parametrization of $g$ is required. A natural and flexible choice for this parametrization is to use neural networks. Neural networks are known to be universal function approximators within certain architectural classes~\cite{Hornik:1989yye, Cybenko:1989iql}, offering the potential to discover effective control variates from data without relying solely on analytical insight or hand-crafted ans\"atze.

In many physical theories, both the action and the observables exhibit translational invariance. To respect this symmetry in the construction of $f$, translational covariance can be enforced in $g$. A practical method for doing so is to define $g$ in terms of a scalar function $g_0:M^V \to \mathbb R$, applied to translated field configurations~\cite{Bedaque:2023ovz}:
\begin{equation}
    g(\phi)_x = g_0(T_x(\phi)) ,
\end{equation}
where $T_x$ denotes the translation operator by a displacement vector $x$, and $g(\phi)_x$ is the $x$th component of the vector-valued function $g$. This construction ensures that $g$ transforms covariantly under lattice translations, thereby preserving the symmetry of the system.

Additional symmetries, such as $Z_2$ invariance under field reflection ($\phi \to -\phi$), can also be imposed on the network. This can be achieved by using zero biases in each layer and selecting odd activation functions. In this work, when $Z_2$ symmetry is required, we use the odd activation function $\sigma(x) \equiv {\rm arcsinh}(x)$.

To train the neural network, we minimize the following loss function:
\begin{equation}
    L(w, \mu) =  \langle ( O-f-\mu)^2\rangle ,
    \label{Eq:Loss}
\end{equation}
where $w$ denotes the parameters of the network and $\mu$ is a learned offset. This loss is preferred over directly minimizing the variance of $O-f$ because it helps prevent overfitting~\cite{Wan2018, South2018, Si2020}.

\section{Correlated configurations} \label{Sec:Correlated}

Since the estimation of statistical errors for observables requires uncorrelated data, it is standard practice to use decorrelated configurations when computing final estimates\footnote{The basics of Markov Chain Monte Carlo and the estimation of errors for correlated samples are summarized in Appendix~\ref{Sec:MCMC}.}. This methodology has been carried over to recent developments in the construction of control variates using Monte Carlo samples, where it is widely assumed that independent configurations are employed for training~\cite{Bhattacharya:2023pxx, Bedaque:2023ovz, Lawrence:2023cft}. The rationale is straightforward: autocorrelated configurations contain no genuinely new information compared to decorrelated ones, so training on them should, in principle, yield similar results to training on a smaller set of independent samples.

However, while correlated samples are generally regarded as statistically redundant for the purposes of error estimation, they are far from meaningless. In fact, correlations arise precisely because the Markov chain explores the probability distribution through local updates. As a result, successive configurations reflect smooth transitions through the configuration space, which encode structural information about the distribution beyond what is captured by a sparse set of decorrelated samples\footnote{The same reasoning applies even when global updates are used in the Markov chain.}. This additional information---arising from correlations between successive configurations in the Markov chain---can be valuable when training control variates or machine learning models. By utilizing correlated samples, one can potentially capture more detailed features of the observable's fluctuation patterns, leading to more effective variance reduction for a given computational budget.

Furthermore, it is important to recognize that the autocorrelation properties of the original observable $O$ may differ from those of the improved observable $\tilde O$ obtained via control variates. While the design of control variates often aims to produce an improved estimator that fluctuates in tandem with $O$, it does not guarantee that the autocorrelation time will be preserved. If $\tilde O$ happens to decorrelate more rapidly than $O$, then discarding correlated configurations---based on the autocorrelation time of $O$---may be counterproductive. In such a scenario, the data considered ``correlated'' for $O$ may in fact be decorrelated, or at least weakly correlated, for $\tilde O$, leading to the unnecessary exclusion of valuable training data. This misalignment underscores the importance of evaluating autocorrelation properties separately for each observable involved in the control variate construction.

In principle, a good control variate should exhibit fluctuations similar to those of $O$, differing only in that its expectation value is zero. This similarity suggests that $f$ might retain comparable autocorrelation properties. Nevertheless, the relationship does not guarantee that $\tilde O$ has the same autocorrelation behavior. Consider, for instance, the perfect control variate defined by $f_P = O - \langle O \rangle$. This choice eliminates all stochastic fluctuations, yielding an estimator with zero variance and, consequently, zero autocorrelation time. While such a perfect control variate is rarely attainable, it highlights that variance reduction can significantly alter the autocorrelation structure, and this should be considered when interpreting results.

Given these considerations, a safer and more flexible strategy is to retain and utilize correlated samples when training control variates. By doing so, one avoids discarding potentially informative data and makes more efficient use of available computational resources---especially in scenarios where generating additional independent configurations is expensive. Importantly, this approach does not conflict with standard practices for error estimation. While training may proceed with correlated data, the final evaluation of observables and their uncertainties should still rely on decorrelated configurations or techniques such as the blocked jackknife or autocorrelation-aware estimators. In this way, one can exploit the full informational content of MCMC samples while maintaining statistical rigor in uncertainty quantification.

\section{Results} \label{Sec:Result}

In this section, we use $U(1)$ gauge theory and scalar field theory to show the numerical evidence of the discussion in the previous section.

\subsection{$U(1)$ gauge theory}

We begin our numerical experiments with a simplified model: two-dimensional $U(1)$ gauge theory with open boundary conditions. Although gauge theories are typically formulated on a lattice in terms of link variables, in two dimensions with open boundaries, it is possible to fix the gauge such that the path integral is written directly in terms of plaquette variables~\cite{Detmold:2020ncp}. The Euclidean action in this formulation is
\begin{equation}
    S = - \beta \sum_x \left(1 - \cos \theta^P_x \right) ,
\end{equation}
and the corresponding partition function becomes
\begin{equation}
    Z = \int \prod_x d\theta^P_x \, {\rm e}^{-\beta \sum_x \left( 1- \cos \theta^P_x \right)} .
\end{equation}

\begin{figure}[t]
    \centering
    \includegraphics[width=0.49\textwidth]{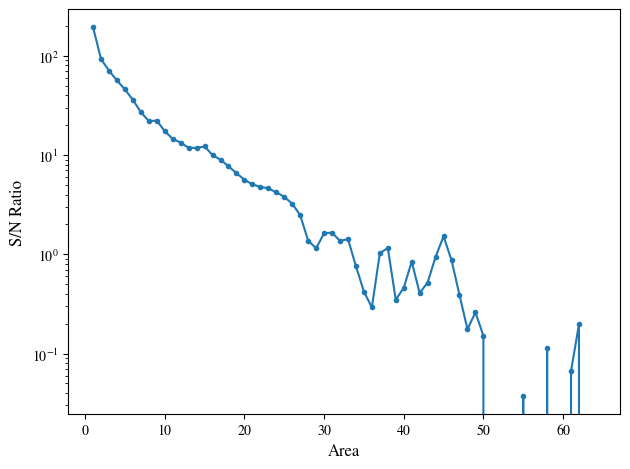}
    
    \caption{Signal-to-noise ratio of Wilson loops with varying areas at coupling $\beta=5.55$, estimated using $10^3$ decorrelated samples.
    }

    \label{fig:u1_StN}
\end{figure}

As our observable, we consider the Wilson loop over a region $A$, which is known to have a signal-to-noise problem:
\begin{equation}
    O(A) = \prod_{x\in A} {\rm e}^{i\theta_p}.
\end{equation}
Throughout this study, we fix the coupling at $\beta=5.55$, corresponding to a fine lattice spacing near the continuum limit~\cite{Detmold:2020ncp}. Figure~\ref{fig:u1_StN} demonstrates that this observable has an exponential signal-to-noise problem by changing the area.

\begin{figure*}[t]
    \centering
    \includegraphics[width=0.49\textwidth]{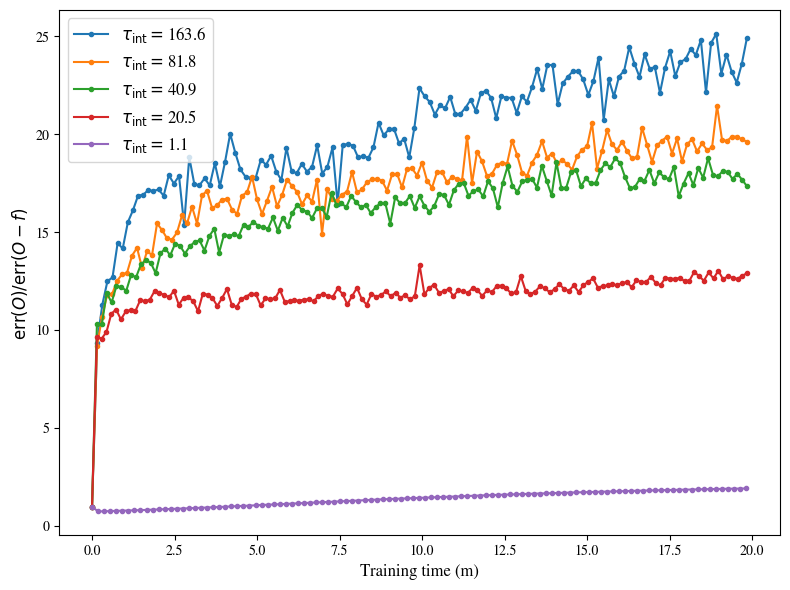}
    \includegraphics[width=0.49\textwidth]{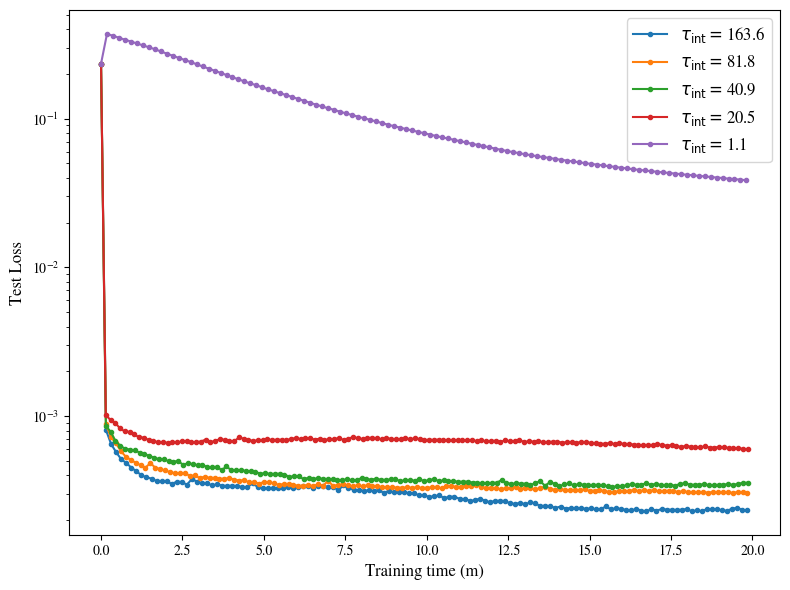}
    
    \caption{Training history for the Wilson loop with area $A=4$ at coupling $\beta=5.55$, using the same Monte Carlo chain samples with different integrated autocorrelation times. The left panel shows the error reduction ratio relative to the original observable $O$. The right panel displays the evolution of the loss function evaluated on decorrelated test configurations.
    }

    \label{fig:u1_4}
\end{figure*}

While Ref.~\cite{Oh:2025jaf} demonstrated that the perfect control variate can be constructed for this model using a specific ansatz and exploiting the factorization properties of both the action and the observable, we deliberately use the general construction given by Eq.~(\ref{Eq:VCV}) to investigate how training with correlated configurations affects the performance of neural control variates.

To isolate this effect, we use unusually large acceptance rates for Metropolis-Hastings updates, ranging from 0.9 to 0.95, significantly higher than typical values (0.3–0.7). Training is performed using $10^4$ correlated Monte Carlo samples, while testing uses $10^3$ fully decorrelated samples obtained by increasing the separation between successive configurations in the Markov chain.

\begin{figure}[t]
    \centering
    \includegraphics[width=0.49\textwidth]{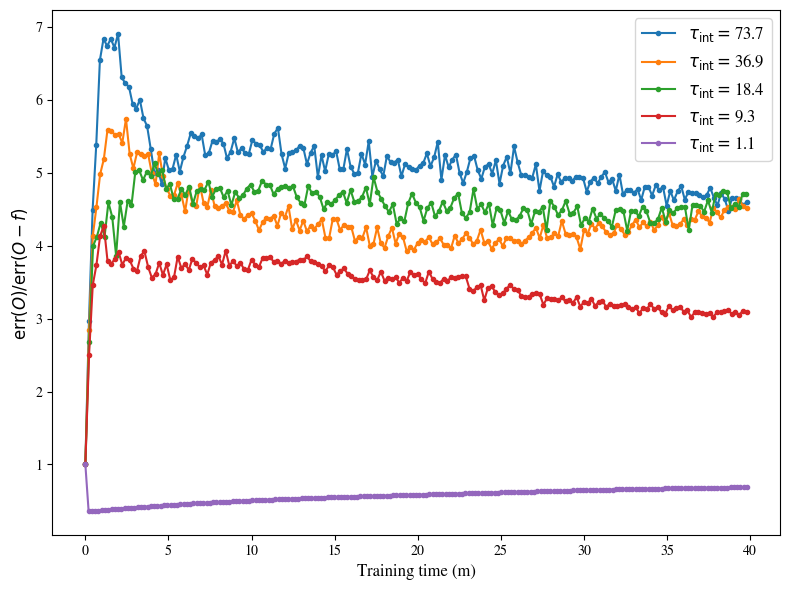}
    
    \caption{Training history for the Wilson loop with area $A=16$ at coupling $\beta=5.55$, using the same Monte Carlo chain sampled with different integrated autocorrelation times.
    }

    \label{fig:u1_16}
\end{figure}

In this toy model, we do not impose any symmetry constraints on the neural network. That is, the network does not implement translational or $Z_2$ symmetries, as discussed in Section~\ref{Sec:NCV}. Instead, we use a generic feedforward network to parametrize $g$, with input features given by $\cos\theta^P$ and $\sin\theta^P$, i.e., the real and imaginary parts of the plaquettes. For the activation function, we choose the Continuously Differentiable Exponential Linear Unit (CELU)~\cite{barron2017}, defined as
\begin{equation}
    {\rm CELU}(x, \alpha) =
    \begin{cases}
        x & \text{if $x\ge0$,} \\
        \alpha \left( \exp \left(x/\alpha \right)-1 \right) & \text{otherwise,}
    \end{cases}
\end{equation}
with $\alpha=1.0$ throughout this work.

The neural network is trained using the \textit{ADAM} optimizer~\cite{kingma2014} with a fixed learning rate of $10^{-4}$ and a minibatch size of 32. To mitigate the risk of exploding gradients, we apply gradient clipping~\cite{Pascanu2013} with a threshold of 1.0.

Figure~\ref{fig:u1_4} shows the impact of correlated samples in training. The left panel presents the variance reduction achieved by the neural control variate relative to the original observable $O$, while the right panel tracks the behavior of the loss function, Eq.~(\ref{Eq:Loss}), during training. The network architecture consists of three hidden layers, each with 32 nodes. We vary integrated autocorrelation times\footnote{It is defined in Appendix~\ref{Sec:MCMC}.} to adjust the level of correlation between training samples and the result for fully decorrelated samples are shown for comparison. This demonstrates that for a fixed length of Monte Carlo chain, more samples while being highly correlated can achieve a better training result than decorrelated samples while they are statistically equivalent for estimating expectation values.

To explore scaling, we increase the area of the Wilson loop to $A=16$, and the corresponding results are shown in Figure~\ref{fig:u1_16}. Here, we use a smaller network with two hidden layers of 16 nodes each. The trend remains consistent: including more correlated configurations in the training data tends to improve variance reduction, given the same total length of the Monte Carlo chain. However, because the loss landscape becomes more complex, which depends on a larger number of variables, the overall variance reduction is less noticeable than in the $A=4$ case.

While the cases with $A=4$ and $A=16$ exhibit less severe signal-to-noise problems compared to larger areas, training neural networks in this context still makes it difficult to clearly distinguish the effect of correlated versus decorrelated training sets, unless large datasets, sufficient training time, and carefully fine-tuned network architectures are used. For this reason, we have focused on presenting results for $A=4$ and $A=16$.

\subsection{Scalar field theory}

\begin{figure}[t]
    \centering
    \includegraphics[width=0.49\textwidth]{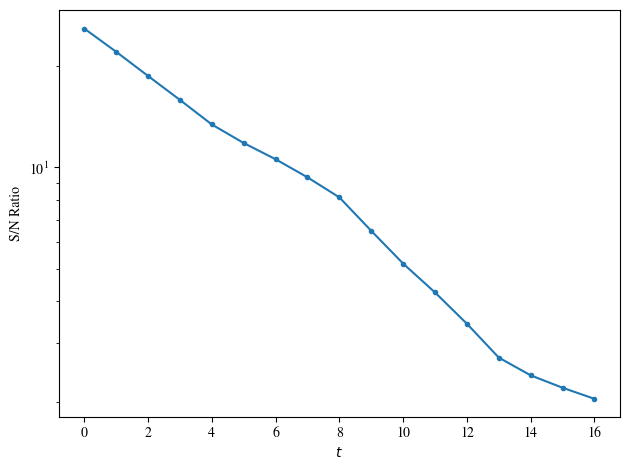}
    
    \caption{Signal-to-noise ratio of the two-point correlator with changing time separation for three-dimensional scalar field theory on a $32 \times 8 \times 8$ lattice with $m^2=0.01$ and $\lambda=0.1$. 
    }

    \label{fig:Scalar_StN}
\end{figure}

\begin{figure*}[t]
    \centering
    \includegraphics[width=0.49\textwidth]{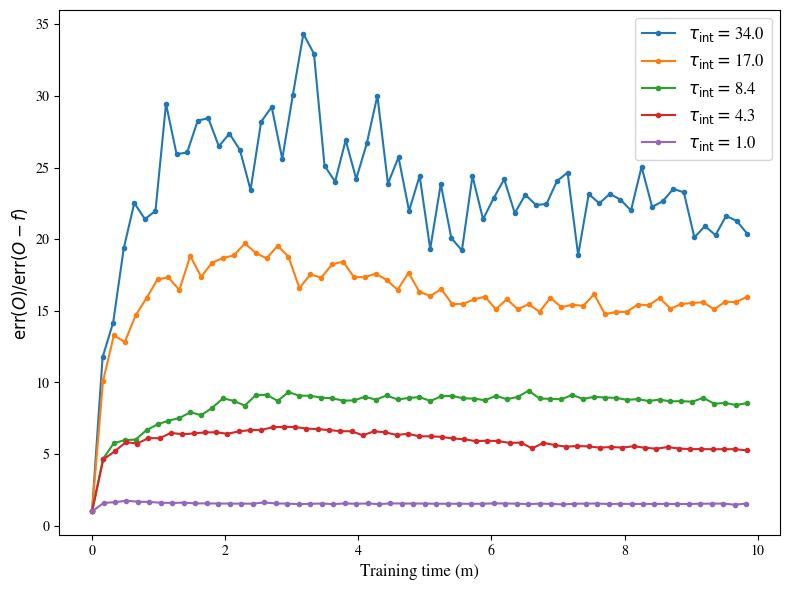}
    \includegraphics[width=0.49\textwidth]{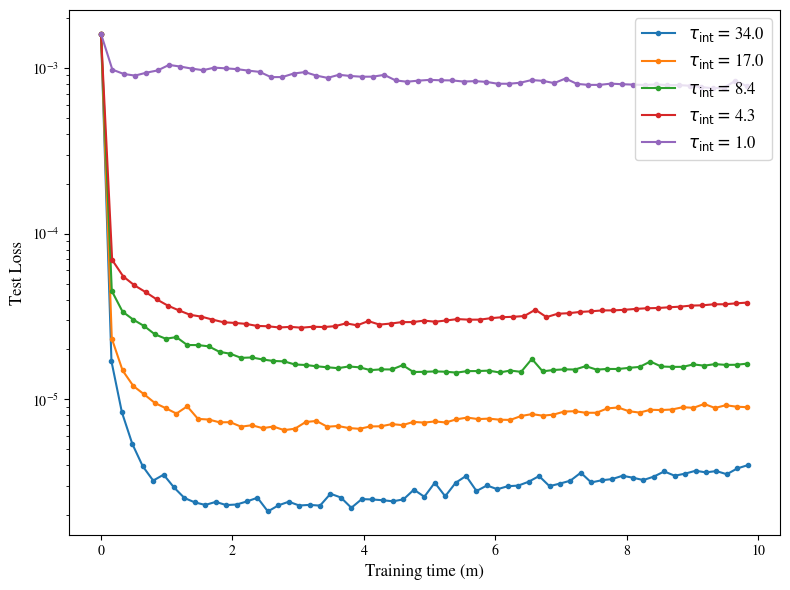}
    \caption{Training history of two-dimensional scalar field theory on a $16 \times 16$ lattice with parameters $m^2=0.1$ and $\lambda=0.5$, using the same Monte Carlo chain sampled with integrated autocorrelation times. The left panel shows the error reduction ratio relative to the original observable $O$. The right panel illustrates the loss function evaluated on decorrelated test configurations throughout training.
    }

    \label{fig:scalar2d}
\end{figure*}

Next, we turn to the scalar field theory with a $\phi^4$ interaction, a well-suited toy model in the context of variance reduction~\cite{Bhattacharya:2023pxx, Bedaque:2023ovz, Lawrence:2022afv, Catumba:2025ljd}. We investigate this theory in two, three, and four spacetime dimensions. The lattice action is given by
\begin{equation}
    S = \frac{1}{2} \sum_{x,\mu} \left(  \phi_{x+\mu} - \phi_x \right)^2 + \frac{m^2}{2} \sum_x \phi_x^4 + \frac{\lambda}{4!}\sum_x \phi_x^4,
\end{equation}
where $ x= (t, \vec{x})$, and $\mu=1,\dots, d$.

The observable considered is the zero-momentum two-point correlation function:
\begin{equation}
    O(t) = \sum_{t'} \left( \frac1V \sum_{\vec{x}} \phi_{t+t', \vec{x}} \right) \left( \frac1V \sum_{\vec{x}} \phi_{t', \vec{x}} \right),
\end{equation}
where $V$ is the spatial volume. Figure~\ref{fig:Scalar_StN} shows that this observable in three dimensions also exhibits an exponential signal-to-noise problem, similar to the Wilson loop. Since the behavior is qualitatively the same in other dimensions, only the three-dimensional case is presented here. In this subsection, we use $t$ as $L_0/2$ on a $L_0 \times L_1 \times \cdots$ lattice, where the signal-to-noise ratio is the worst.

Monte Carlo configurations are generated using the Metropolis-Hastings algorithm, with measurements performed at every sweep. The training set consists of $10^4$ samples, while the test set comprises $10^3$ fully decorrelated configurations obtained by inserting long intervals between measurements.

As the $U(1)$ gauge theory case, we use the \textit{ADAM} optimizer with a fixed learning rate of $10^{-4}$, a minibatch size of 32, and gradient clipping with a threshold of 1.0. For simplicity, $L_2$ regularization, which was utilized in~\cite{Bedaque:2023ovz}, is not used. Since our focus is on understanding the role of correlated samples in training, regularization---which primarily addresses overfitting---is not expected to qualitatively affect the results.

Figure~\ref{fig:scalar2d} shows the results in two dimensions. The left panel displays the error reduction relative to the original observable $O$, while the right panel shows the evolution of the loss function, Eq.~(\ref{Eq:Loss}), during training. The neural network used has two hidden layers with 32 nodes each. Notably, the loss function begins to rise after a certain number of training steps, indicating the need for early stopping.


\begin{figure*}[t]
    \centering
    \includegraphics[width=0.49\textwidth]{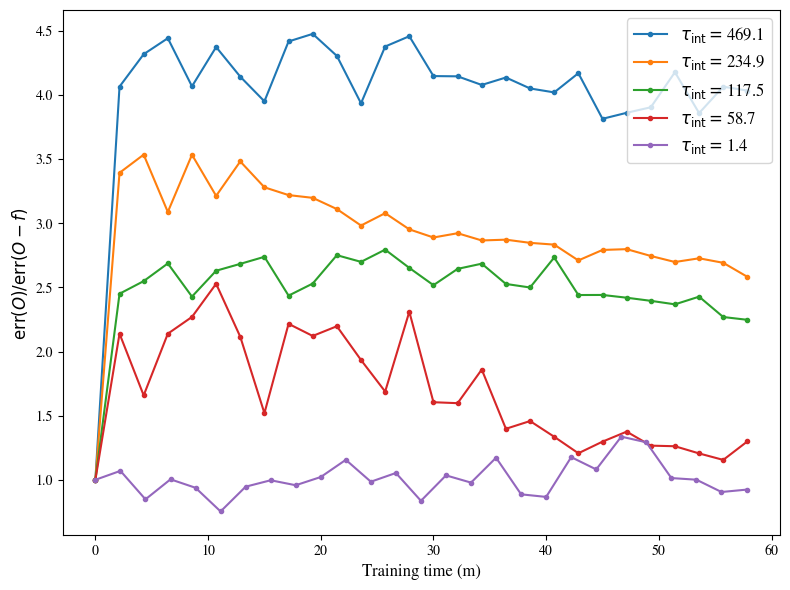}
    \includegraphics[width=0.49\textwidth]{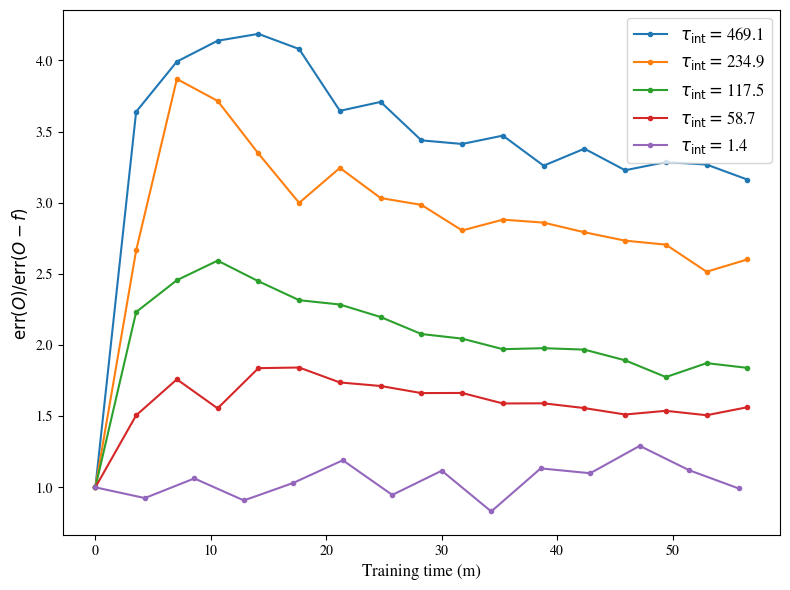}
    
    \caption{Training history of three-dimensional scalar field theory on a $32 \times 8 \times 8$ lattice with parameters $m^2=0.01$ and $\lambda=0.1$, using the same Monte Carlo chain sampled with integrated autocorrelation times.
    The left panel shows the error reduction ratio relative to the original observable $O$, using a neural network with two hidden layers of 32 nodes each. The right panel shows the same quantity using a deeper network with four hidden layers of 64 nodes each.
    }

    \label{fig:scalar3d}
\end{figure*}

Figure~\ref{fig:scalar3d} presents results in three dimensions on a $32 \times 8 \times 8$ lattice. Two network architectures are compared: the left panel uses two hidden layers with 32 nodes, and the right panel uses four hidden layers with 64 nodes. In both cases, error reduction improves with more training samples, even when those samples are correlated.

\begin{figure}[t]
    \centering
    \includegraphics[width=0.49\textwidth]{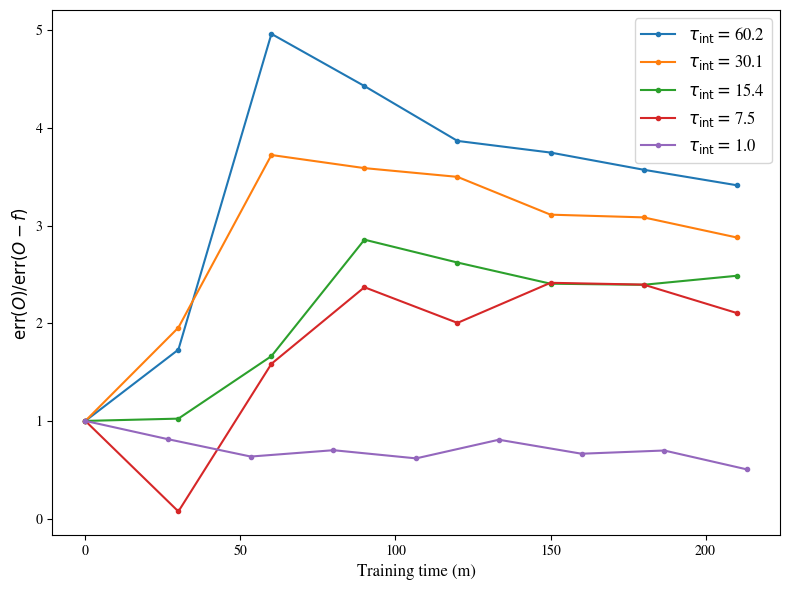}
    
    \caption{
    Training history of four-dimensional scalar field theory on a $16 \times 8 \times 8 \times 8$ lattice with parameters $m^2=0.1$ and $\lambda=0.5$, using the same Monte Carlo chain sampled with different measurement intervals (in units of sweeps).
    }

    \label{fig:scalar4d}
\end{figure}

Finally, Figure~\ref{fig:scalar4d} shows results in four dimensions. As in the previous cases, the figure demonstrates that incorporating a larger number of samples---even if correlated---leads to better variance reduction, given a fixed MCMC chain and training budget.

\section{Discussion} \label{Sec:Discussion}

In this work, we demonstrated that using correlated samples to train neural control variates can lead to more effective variance reduction, even when the total length of the Monte Carlo chain is fixed. Through studies of scalar field theory and $U(1)$ gauge theory, we showed that configurations typically discarded for error estimation due to autocorrelation can, in fact, be valuable for improving the training of neural control variates.

Our results highlight that correlated data---often ignored in standard workflows---can serve as a useful resource in constructing these variates. This contributes both conceptual insight and practical guidance for deploying neural control variates in computational physics, particularly in large-scale lattice simulations and other complex systems.

A key consideration for scaling this approach to larger volumes, such as those encountered in state-of-the-art lattice QCD calculations, is the limited availability of independent configurations---typically on the order of $\sim 10^3$ from large-scale simulations on high-performance computing resources. While the number of independent samples may be small, our findings suggest that the many correlated configurations generated in between can be repurposed to train better control variates.

Nonetheless, storing every configuration after each Monte Carlo update is computationally expensive in terms of memory. Moreover, consecutive configurations may be so similar that their contribution to learning an effective control variate is minimal. In this study, we did not systematically investigate the optimal measurement frequency for training, which remains an important direction for future research.

Finally, our findings may extend to other neural network-based methods that rely on Monte Carlo sampling. Techniques such as contour deformation~\cite{Alexandru:2020wrj}, neural network quantum states~\cite{Carleo2017}, and neural network assisted Monte Carlo algorithms~\cite{Liu:2017muj, levy2018, Muller2019} could potentially benefit from incorporating correlated samples into their training procedures. Whether this approach improves those methods remains an open question for future exploration.

\begin{acknowledgments}

This work was supported in part by the U.S. Department of Energy, Office of Nuclear Physics under Award Number DE-FG02-93ER40762. 

\end{acknowledgments}

\bibliography{refs.bib}

\appendix

\section{Markov Chain Monte Carlo and error estimation} \label{Sec:MCMC}

\begin{figure*}[t]
    \centering
    \includegraphics[width=0.49\textwidth]{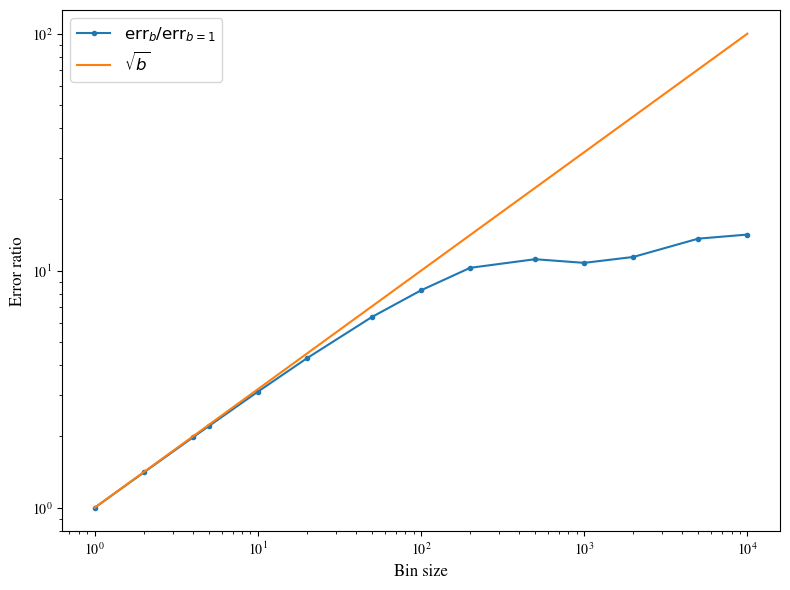}
    \includegraphics[width=0.49\textwidth]{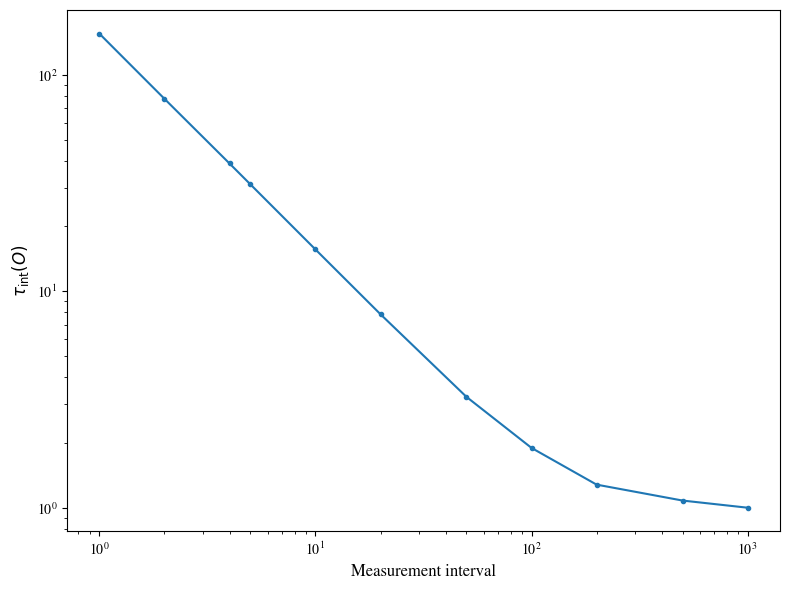}
    
    \caption{An example of estimating the autocorrelation time from $10^5$ correlated configurations. The left panel shows the ratio of jackknife errors at each bin size to the error at bin size 1. The right panel illustrates the integrated autocorrelation time with respect to different measurement intervals.
    }

    \label{fig:correlated}
\end{figure*}

Markov Chain Monte Carlo (MCMC) methods are a family of algorithms used to approximate complex probability distributions through iterative sampling. These techniques are especially valuable when direct sampling or analytical integration is infeasible. MCMC generates samples by constructing a Markov chain that has the desired distribution as its equilibrium distribution. After an initial burn-in period, the chain converges, and subsequent samples can be used to estimate expectations, variances, or other characteristics of the target distribution. Common algorithms within this framework include the Metropolis-Hastings algorithm~\cite{Metropolis:1953am, Hastings:1970aa} and the Gibbs sampler~\cite{Geman:1984wrl}.

A critical consideration when using MCMC is the presence of autocorrelation between successive samples. Since each new sample depends on the previous one, the resulting sequence can exhibit significant correlation, especially if the chain mixes slowly. This autocorrelation reduces the effective number of independent samples, which can impair the accuracy of statistical estimates derived from the chain.

In order to estimate the error of an observable from a finite sample, the jackknife method~\cite{Quenouille1949, Quenouille1956} is widely used. It is a resampling technique used to estimate the variance of a finite sample. For a data set $\{x_1, x_2, \dots, x_n\}$, let $\hat\theta$ be the estimator of interest computed using the full dataset. The jackknife estimate is constructed by systematically omitting one observable at a time. Define the $i$th jackknife replicate as
\begin{equation}
    \hat{\theta}_{(i)} = \hat{\theta}(x_1, \dots, x_{i-1}, x_{i+1}, \dots, x_n).
\end{equation}
Then the jackknife estimate of the variance is given by
\begin{equation}
    {\rm Var}_{\rm jack}(\hat{\theta}) = \frac{n-1}{n} \sum_{i=1}^n \left( \hat{\theta}_{(i)} - \bar{\theta}\right)^2,
\end{equation}
where $\bar{\theta}$ is the sample average of the estimator $\hat\theta$. The error (standard deviation) becomes
\begin{equation}
    {\rm err}_{\rm jack}(\hat\theta) =  \sqrt{ {\rm Var}_{\rm jack}(\hat\theta)}\footnote{For a population mean, $\hat \theta = \frac{1}{n} \sum_{i=1} \theta_i$, it implies the famous formula, ${\rm err}_{\rm jack}(\hat\theta) = \sqrt{\frac{1}{n} {\rm Var}(\theta) }$.}.
\end{equation}
This method assumes that data points are independent. When this assumption is violated, such as correlated samples from MCMC, a modified version of the jackknife is required.

To address the autocorrelation of the data, the jackknife with bin size (also known as the blocked jackknife) groups the data into $B$ non-overlapping bins, each of size $b$, such that $n=B \cdot b$. Each bin $j$ consists of $\{ x_{(j-1)b+1}, \dots, x_{jb} \}$. The $j$th jackknife replicate is computed by omitting the entire $j$th bin:
\begin{equation}
    \hat{\theta}_{(j)} = \hat\theta(\text{data without bin }j).
\end{equation}
The variance estimate becomes
\begin{equation}
    {\rm Var}_{\text{jack}, b} (\hat\theta) = \frac{B-1}{B}\sum_{j=1}^B \left( \hat{\theta}_{(j)} - \bar\theta\right)^2.
\end{equation}
By choosing an appropriate bin size $b$, this method accounts for within-bin correlation and provides more reliable estimates. In practice the bin size is often chosen by checking the plateau of the size of the variance by increasing the size of the bin size.

Another common way to characterize autocorrelation in a Markov chain is through the integrated autocorrelation time, $\tau_{\rm int}$. This quantity is defined by
\begin{equation}
    \tau_{\rm int}(\theta) \equiv 1+2\sum_{i=1}^{\infty} \frac{{\rm Cov}(\theta_k, \theta_{k+i})}{\sigma^2},
\end{equation}
where $\sigma^2= {\rm Var}(\theta)$ is the variance of the observable. The integrated autocorrelation time measures how strongly each sample is correlated with its successors and quantifies how many successive samples are effectively equivalent to a single independent measurement. As a result, for an estimator $\bar{\theta}$ based on $n$ samples, the variance is inflated by $\tau_{\rm int}$:
\begin{equation}
    {\rm Var}(\bar{\theta}) = \frac{\sigma^2}{n} \tau_{\rm int}.
\end{equation}
In other words, autocorrelation reduces the effective number of independent samples by a factor of $\tau_{\rm int}$, which must be accounted for when estimating statistical uncertainties.

The left panel of Figure~\ref{fig:correlated} presents the error estimation of correlated samples using the blocked jackknife method with varying bin sizes. When the bin size is small and insufficient to capture the full autocorrelation of the data, the estimated error increases approximately as $\sqrt{b}$, where $b$ is the bin size. This behavior arises because the underlying correlations between samples are not adequately removed, causing the jackknife estimator to underestimate the true variance at small bin sizes.

Similarly, the right panel of Figure~\ref{fig:correlated} shows how the integrated autocorrelation time changes with increasing measurement interval. As the interval between measurements grows, the samples become progressively less correlated, and the autocorrelation time approaches 1, indicating fully independent configurations.

\end{document}